\documentclass[pss]{wiley2sp} % provides pss two-column style
\usepackage{amsmath}
\usepackage{bm}
\newcommand{\rmi}{{\rm i}}
\newcommand{\ve}[1]{{\bm #1 }}
\newcommand{\hc}{\hat{c}^{\phantom{\dagger}}}
\newcommand{\hcd}{\hat{c}^{\dagger}}
\newcommand{\ket}[1]{\left| #1 \right \rangle }  
\newcommand{\bra}[1]{\left \langle #1 \right |} 
\newcommand{\rangind}[1]{\rangle\hspace{-0.2cm}{\phantom{\rangle}}_{#1}
\hspace{0.05cm}}
\newcommand{\langind}[1]{\hspace{-0.1cm}{\phantom{\rangle}}_{#1}
\hspace{-0.05cm} \langle\hspace{0.03cm} }  

\begin{document}
\title{Numerical Minimisation of Gutzwiller Energy Functionals}
% Title of the article

% Abbreviated title for the page headers
\titlerunning{Minimisation of Gutzwiller Energy Functionals}

% Authors
\author{J\"org B\"unemann\textsuperscript{\Ast,\textsf{\bfseries 1}}, 
Florian Gebhard\textsuperscript{\textsf{\bfseries 2}}, 
Tobias Schickling\textsuperscript{\textsf{\bfseries 2}}, and 
Werner Weber\textsuperscript{\textsf{\bfseries 3}}}

\authorrunning{J.\ B\"unemann et al.}
\mail{e-mail
  \textsf{buenemann@gmail.com}}

\institute{%
  \textsuperscript{1}\,Institut f\"ur Physik, BTU Cottbus, 
P.O.\ Box 101344, 03013 Cottbus, Germany\\
  \textsuperscript{2}\,Fachbereich Physik, Philipps Universit\"at, 
Renthof 6, 35032 Marburg, Germany\\
  \textsuperscript{3}\,Theoretische Physik II,
Technische Universit\"at Dortmund, Otto-Hahn-Str.\ 4,
44227 Dortmund, Germany}

\received{XXXX, revised XXXX, accepted XXXX} % do not change, will be filled in by the publisher
\published{XXXX} % do not change, will be filled in by the publisher

% Please select about four verbal keywords for your manuscript.

\abstract{
\abstcol{We give a comprehensive introduction into an efficient 
numerical scheme for the 
 minimisation of Gutzwiller energy functionals for
multi-band Hubbard models.}
{Our method covers all conceivable
cases of Gutzwiller variational  wave functions and has been used successfully 
in previous numerical studies. }
}

\maketitle   % please do not remove

\section{Introduction}

In solid-state theory, multi-band Hubbard models
are used to study transition metals and 
their compounds. In these models only 
the local (atomic) 
  part of the Coulomb interaction is explicitly taken into account. All 
non-local terms are included on the level of a `Density-Functional 
 Theory' calculation, which is used to set up a proper tight-binding 
Hamiltonian, see Sect.~\ref{sec1}. 

Despite the relative simplicity of 
 Hubbard models, as compared to the full electronic Hamiltonian, calculating 
 their properties still constitutes a very difficult many-particle 
 problem. In recent years, significant progress has been made 
in this direction by the systematic study of models in
 the limit of infinite spatial dimensions 
($D\to \infty$). The exact solution of Hubbard models in this limit leads 
to the Dynamical Mean Field Theory (DMFT), in which the original lattice model 
is mapped onto an effective single-impurity system that has to be solved 
numerically \cite{metzner1989,vollhardt1989,vollhardt1993,georges1996,gebhard1997}. Although significant progress has been made in recent years 
in developing numerical techniques for the solution of the DMFT equations, 
it is still quite challenging  and can be carried out only with 
limited  accuracy. 

An alternative method, that also relies on infinite-$D$ techniques, is the 
Gutzwiller 
 variational approach. It allows for the approximate study of ground-state
 properties and single-particle excitations
 with much less numerical
 effort than within DMFT and has been applied in a number of works in 
 recent years \cite{buenemann1997c,buenemann1998,buenemann2003,attaccalite2003,buenemann2003b,ferrero2005,julien2005,buenemann2005,buenemann2007b,buenemann2007d,buenemann2008,lanata2008,ho2008,deng2009,zhouang2009,borghi2009,hofmann2009,wang2010,zhou2010,buenemann2011,yao2011,buenemann2011e}. A related approach that leads to the
 same energy functional for multi-band models is the slave-boson 
mean field  theory 
\cite{lechermann2007,buenemann2007c,ferrero2009,isidori2009,lechermann2009,buenemann2010,piefke2011}. Starting from the approximate ground-state description, it is
 also possible to study two-particle excitations  within the 
`time-dependent Gutzwiller
 theory' \cite{seibold1998b,seibold2003,lorenzana2003,seibold2004,seibold2004b,lorenzana2005,seibold2005,seibold2006,seibold2007,seibold2008,seibold2008b,guenther2010,buenemann2011b,buenemann2011c}.
 
The main numerical problem in  the Gutzwiller theory is 
the minimisation of the energy functional with respect to the variational 
parameters since their number can be quite large in investigations
 of multi-band models. 
 We have developed an efficient numerical scheme
 for this minimisation which has already been applied successfully in our 
 studies on nickel \cite{buenemann2003,buenemann2008} and iron-pnictides 
\cite{buenemann2011,buenemann2011e}. In particular, the studies on the 
 spin-orbit coupling effects in nickel were numerically demanding 
 since they required a rather fine energy resolution and the handling of up 
to 8000 variational parameters \cite{buenemann2008}. To the best of our 
knowledge, no Gutzwiller minimisation of 
similar complexity has been reported in other works.
 We are therefore convinced that our minimisation algorithm will be of 
significant interest for all researchers who intend to apply the 
Gutzwiller theory to real  materials. It is the purpose 
 of this work to give detailed account of our method. Note that an alternative 
method for the minimisation of a restricted class 
of Gutzwiller energy functionals has been proposed in a  
recent work~\cite{lanata2011}. 

Our presentation is organised as follows. In Sections~\ref{sec1}  
 and~\ref{chap2b} we summarise the main results on multi-band 
 Gutz\-willer wave functions and their energy functionals in infinite 
 spatial dimensions. Our minimisation algorithm is described in detail 
 in  Section~\ref{app7}. Some technical parts of the presentation
 are referred to four appendices.  
 
\section{Multi-Band Hubbard models}\label{sec1}

We aim to study the physics 
of  multi-band Hubbard models 
\begin{equation}\label{h2}
\hat{H}=\sum_{i\neq j} \sum_{\sigma,\sigma'}
t^{\sigma,\sigma'}_{i,j} \hcd_{i,\sigma}\hc_{j,\sigma'}+
\sum_i \hat{H}_{i,{\rm loc}}\;.
\end{equation}
Here, we introduced the `hopping parameters' $t^{\sigma,\sigma'}_{i,j}$ and 
 the operators $\hat{c}^{(\dagger)}_{i,\sigma}$, which annihilate (create)
 an electron with spin-orbital index $\sigma$ on a lattice site $i$. 
The  local Hamiltonian 
\begin{eqnarray}\label{4.10a}
\hat{H}_{i;{\rm loc}}&=&\sum_{\sigma_1,\sigma_2}\varepsilon_{i;\sigma_1,\sigma_2}
\hcd_{i,\sigma_1} \hc_{i,\sigma_2}\\ \nonumber
&&+\sum_{\sigma_1,\sigma_2,\sigma_3,\sigma_4}
U_i^{\sigma_1,\sigma_2,\sigma_3,\sigma_4}
\hcd_{i,\sigma_1} \hcd_{i,\sigma_2}\hc_{i,\sigma_3} \hc_{i,\sigma_4}\;
\end{eqnarray}
is determined  by the orbital-dependent on-site energies  
$\varepsilon_{i;\sigma_1,\sigma_2}$
 and by the two-particle Coulomb interaction matrix elements
$U_i^{\sigma_1,\sigma_2,\sigma_3,\sigma_4}$.
 We assume that the $2N$ spin-orbital
 states $\sigma$ are  ordered in some arbitrary 
 way, $\sigma= 1,\ldots,2N$ where $N$ is the number of orbitals
 per lattice site.
In order to set up a proper basis of the local Hilbert space, 
 we introduce the following notations for the $2^{2N}$ possible 
configurations.
\begin{itemize}
\item[i)\,] An atomic configuration $I$ is characterised by the electron 
occupation of the orbitals,
\begin{eqnarray}\label{4.20a}
 I &\in&  \{\emptyset;
(1),\ldots,(2N);
(1,2),\ldots,(2,3),\\\nonumber
&&\ldots (2N-1,2N);
\ldots;
(1,\ldots,2N)
\}\;,
\end{eqnarray}
where the elements in each set $I=(\sigma_1,\sigma_2,\ldots)$ 
are ordered, i.e., it is $\sigma_1<\sigma_2<\ldots$. The symbol
$\emptyset$ in~(\ref{4.20a}) means that the site is empty.
In general, we interpret the indices $I$ as sets in the usual 
 mathematical sense.
For example, in the atomic configuration
$I\backslash I'$ 
only those orbitals in $I$ that are not in $I'$ 
are occupied.
The complement of $I$ is 
$\overline{I}\equiv(1,2,\ldots,2N)\backslash I$,
i.e., in the atomic configuration $\overline{I}$ all orbitals but those
in $I$ are occupied.
%\newline 
\item[ii)\,] The absolute value $|I|$ of a configuration 
is the number of elements in it, i.e.,
\begin{eqnarray}\label{4.25a}
&&|\emptyset|=0;|(\sigma_1)|=1;|(\sigma_1,\sigma_2)|=2;\\\nonumber
&&\ldots;|(1,\ldots,2N)|=2N
\;.  
\end{eqnarray}
\item[iii)\,] A state with a specific configuration $I$ is given as 
\begin{equation}\label{4.30a}
\ket{I}=\hat{C}_{I}^{\dagger}\ket{0}\equiv\prod_{\sigma \in I}\hcd_{\sigma}\ket{0}=
\hcd_{\sigma_1}\dots\hcd_{\sigma_{|I|}}\ket{0}\;,
\end{equation}
where the operators $\hcd_{\sigma}$ are in ascending order, i.e., it is 
 $\sigma_1<\sigma_2\ldots<\sigma_{|I|}$. 
Products of annihilation operators, such as
\begin{equation}\label{4.35a}
\hat{C}_{I}^{}\equiv\prod_{\sigma\in I}\hc_{\sigma}=\hc_{\sigma_1}\dots\hc_{\sigma_{|I|}},
\end{equation}
will be  placed in descending order, i.e., with 
$\sigma_1>\sigma_2\ldots>\sigma_{|I|}$. Note that we have introduced  the operators
 $\hat{C}_{I}^{\dagger}$ and  $\hat{C}_{I}^{}$ just as convenient abbreviations.
 They must not be misinterpreted as 
  fermionic creation or annihilation operators.
\item[iv)\,] The operator $\hat{m}_{I,I'}\equiv \ket{I}\bra{I'}$ 
describes the transfer
 between configurations $I'$ and $I$. It can be written as  
 \begin{equation}\label{4.50a}
\hat{m}_{I,I'}=
\hat{C}_{I}^{\dagger}
\hat{C}_{I'}^{}
\prod_{\sigma''\in J}(1-\hat{n}_{\sigma''})
\end{equation}
where $J\equiv \overline{I\cup I'}$. A special case,
  which derives from~(\ref{4.50a}), is the occupation operator
\begin{equation}\label{4.52a}
\hat{m}_{I}\equiv  \ket{I}\bra{I}=\prod_{\sigma\in I}\hat{n}_{\sigma}
\prod_{\sigma'\in \bar{I}}(1-\hat{n}_{\sigma'})\;.
\end{equation}
\end{itemize}
The states $\ket{I}$ form a basis of the atomic Hilbert space. Therefore,
 we can write the eigenstates of the local Hamiltonian~(\ref{4.10a}) as
\begin{equation}\label{4.60a}
|\Gamma\rangle =\sum_{I}T_{I,\Gamma}\ket{I}
\end{equation}
with coefficients $T_{I,\Gamma}$. 
With these eigenstates, the atomic Hamiltonian has the form
\begin{eqnarray}\label{eft}
\hat{H}_{i,{\rm loc}}&=&\sum_{\Gamma}
E_{i;\Gamma}\hat{m}_{i;\Gamma,\Gamma}\;,\\
\hat{m}_{i;\Gamma,\Gamma'}&\equiv&
|  \Gamma\rangind{i}
\langind{i} \Gamma'| =
\sum_{I,I'}T_{I,\Gamma}T^*_{I',\Gamma'}|  
I\rangind{i}
\langind{i} I'|
\;.
\end{eqnarray}
\section{Gutzwiller Energy Functional}\label{chap2b}

Multi-band Gutz\-wil\-ler wave-functions have the form
\begin{equation}\label{1.3}
|\Psi_{\rm G}\rangle=\hat{P}_{\rm G}|\Psi_0\rangle=\prod_{i}\hat{P}_{i}|\Psi_0\rangle\;,
\end{equation}
where $|\Psi_0\rangle$ is a normalised single-particle product state and the 
local Gutzwiller correlator is defined as 
\begin{equation}\label{1.4b}
\hat{P}_{i}=\sum_{\Gamma,\Gamma^{\prime}}\lambda_{i;\Gamma,\Gamma^{\prime}}
|\Gamma \rangle_{i} {}_{i}\langle \Gamma^{\prime} |\equiv
 \sum_{\tilde{\Gamma}}\lambda_{i;\tilde{\Gamma}} 
| \tilde{\Gamma} \rangle_{i} {}_{i}\langle \tilde{\Gamma}  |   \;,
\end{equation}
where we introduced the matrix of variational parameters
$\lambda_{i;\Gamma,\Gamma^{\prime}}$ which allows us to optimise the occupation 
 and the form of the eigenstates $|\tilde{\Gamma} \rangle_{i}$ of $\hat{P}_{i}$.

The evaluation of expectations values with respect to the 
 wave function~(\ref{1.3}) is a difficult many-particle problem, 
which cannot be 
 solved in general. 
 As shown 
in~Refs.~\cite{buenemann1998,buenemann2005}, 
one can derive analytical expressions for the 
 variational ground-state energy in the limit of infinite spatial dimensions 
($D\to \infty$). Using this energy functional 
 for the study of finite-dimensional systems  is usually 
denoted as the `Gutzwiller approximation'.  This approach is the basis 
of most applications of Gutzwiller wave functions in studies 
of real materials and it will also be addressed in this work. 
One should keep in mind, however, that the Gutzwiller approximation has its 
limitations and the study of some phenomena requires an evaluation 
 of expectation values in finite dimensions~\cite{buenemann2011d}.

 \subsection{Local basis}
In general, the local density matrix for non-interacting electrons
\begin{equation}\label{xc}
C_{i;\sigma,\sigma'}=\langle \hcd_{i,\sigma}\hc_{i,\sigma'}   \rangle_{\Psi_0}
\end{equation}
 is non-diagonal with respect to $\sigma,\sigma'$.  For a fixed state 
 $\ket{\Psi_0}$, one can always find a local basis with a 
diagonal density matrix. 
 This will turn out to be quite useful in the minimisation with respect to 
 the variational parameters $\lambda_{i;\Gamma,\Gamma^{\prime}}$
 because, with such a basis,  the energy functional has a much simpler form. 
We introduce the explicit expression of this simplified functional 
in the following Sects.~\ref{con1} and~\ref{exp}. 
 If one minimises the energy with respect to $\ket{\Psi_0}$, however, 
 the diagonality of~(\ref{xc}) is only ensured in systems with 
 high symmetries. Therefore, we also need the general expression
 for the variational ground-state energy with an arbitrary local 
 basis. This is given in Appendix~\ref{ka}. 

Note that, in general, the {\sl correlated} density matrix
\begin{equation}\label{xc1}
C^{\rm c}_{i;\sigma,\sigma'}=\langle \hcd_{i,\sigma}\hc_{i,\sigma'}   
\rangle_{\Psi_{\rm G}}
\end{equation}
is different from the {\sl non-interacting} density matrix~(\ref{xc}). 
 In the following, however, we will frequently use the short term 
`density matrix' 
for~(\ref{xc})  since the correlated density 
matrix~(\ref{xc1}) is not considered in this work.  
Moreover, we only study systems and wave functions
 which are translationally invariant. Therefore we drop lattice
 site indices whenever this does not create ambiguities.

\subsection{Constraints}\label{con1}
  As shown in Refs.~\cite{buenemann1998,buenemann2005}, it is most 
 convenient for the evaluation of Gutzwiller wave functions in 
 infinite dimensions to impose the following (local) constraints
\begin{eqnarray}\label{1.10a}
\langle\hat{P}^{\dagger}\hat{P}^{}\rangle_{\Psi_0}&=&1\;,\\
\label{1.10b}
\langle  \hat{c}^{\dagger}_{\sigma} \hat{P}^{\dagger}\hat{P}^{} \
\hat{c}^{}_{\sigma'}\rangle_{\Psi_0}&=&\langle
 \hat{c}^{\dagger}_{\sigma}\hat{c}^{}_{\sigma'}   \rangle_{\Psi_0}\;. 
\end{eqnarray}
Note  that moving the operator $\hat{P}^{\dagger}\hat{P}^{}$  relative to 
$\hat{c}^{\dagger}_{\sigma}$ or $\hat{c}^{}_{\sigma'}$ 
 in (\ref{1.10b})  does not alter the whole 
set of constraints. With the explicit form of the 
 correlation operator~(\ref{1.3}) and an orbital basis with a diagonal 
 local density matrix,
\begin{equation}\label{xc2}
C_{\sigma,\sigma'}=\delta_{\sigma,\sigma'}n_{\sigma}\;,
\end{equation}
 the constraints read as
\begin{eqnarray}\label{5.5}
\sum_{\Gamma,\Gamma_1,\Gamma_2}
\lambda_{\Gamma,\Gamma_1}^{*}\lambda_{\Gamma,\Gamma_2}^{}
m^{0}_{\Gamma_1,\Gamma_2}&=&1\;,\\\label{5.5b}
\sum_{\Gamma,\Gamma_1,\Gamma_2}
\lambda_{\Gamma,\Gamma_1}^{*}\lambda_{\Gamma,\Gamma_2}^{}
m^{0}_{\Gamma_1\cup \sigma,\Gamma_2\cup \sigma'}
&=&\delta_{\sigma,\sigma'}n_{\sigma}\;,
\end{eqnarray}
where  
\begin{eqnarray}\label{wet}
|\Gamma\cup \sigma \rangle
&\equiv& \hat{c}^{\dagger}_{\sigma}|\Gamma \rangle
 =\sum_{I(\sigma \notin I)}T_{I,\Gamma}|I \cup \sigma\rangle\;,\\\label{wet2}
m^{0}_{\Gamma,\Gamma'}&=&\langle \hat{m}_{\Gamma,\Gamma'} \rangle_{\Psi_0}=
\sum_{I}T_{I,\Gamma}T^*_{I,\Gamma'}m^{0}_{I}\;,\\
m^{0}_{I}&=&\prod_{\sigma \in I}n_{\sigma}\prod_{\sigma \notin I}(1-n_{\sigma})  \;.
\end{eqnarray}
For a general orbital basis the explicit form of the constraints is given 
 in Appendix~\ref{ka}.

\subsection{Expectation values}\label{exp}
Each local operator $\hat{O}_i$, e.g., the local Hamiltonian~(\ref{4.10a}), 
can be written as
\begin{equation}
\hat{O}_i=\sum_{\Gamma,\Gamma'}O_{\Gamma,\Gamma'}\hat{m}_{i;\Gamma,\Gamma'}\;.
\end{equation}
In infinite dimensions, its expectation value with respect to~(\ref{1.3}) 
is given as
 \begin{equation}
\langle 
\hat{O}\rangle_{\Psi_{\rm G}}
=\sum_{\Gamma_1,\Gamma_2,\Gamma_3,\Gamma_4}
O_{\Gamma_2,\Gamma_3}
\lambda_{\Gamma_2,\Gamma_1}^{*}\lambda_{\Gamma_3,\Gamma_4}^{}
m^0_{\Gamma_1,\Gamma_4}\;,
\end{equation}
where the expectation values $m^0_{\Gamma,\Gamma'}$ have been introduced 
in~(\ref{wet2}).  
Hence, the expectation value of the local Hamiltonian~(\ref{eft}) becomes
\begin{equation}\label{kdr} 
\langle 
\hat{H}_{i,{\rm loc}}\rangle_{\Psi_{\rm G}}
=\sum_{\Gamma,\Gamma_1,\Gamma_2}E_{\Gamma}
\lambda_{\Gamma,\Gamma_1}^{*}\lambda_{\Gamma,\Gamma_2}^{}
m^0_{\Gamma_1,\Gamma_2}\;.
\end{equation}

The expectation value for a hopping operator in infinite dimensions 
has the form
\begin{equation}\label{8.410} 
\big \langle  \hat{c}_{i,\sigma_1}^{\dagger}\hat{c}_{j,\sigma_2}^{\phantom{+}} \big \rangle_{\Psi_{\rm G}}
=\sum_{\sigma'_1,\sigma'_2}q_{\sigma_1}^{\sigma'_1}\left( q_{\sigma_2}^{\sigma'_2}\right)^{*}\big \langle  
\hat{c}_{i,\sigma'_1}^{\dagger}\hat{c}_{j,\sigma'_2}^{\phantom{+}} \big \rangle_{\Psi_{0}}\;,
\end{equation}
where, for an orbital basis with diagonal local density matrix,
 the (local) renormalisation matrix reads 
\begin{eqnarray}\nonumber
q_{\sigma}^{\sigma'}&=&\frac{1}{n_{\sigma'}}
\sum_{\Gamma_1\ldots\Gamma_4}\lambda^{*}_{\Gamma_2,\Gamma_1}
\lambda^{}_{\Gamma_3,\Gamma_4}
\langle \Gamma_2|
\hcd_{\sigma}
|\Gamma_3\rangle\\\label{8.460} 
 &&\times \Big  \langle
\big (|\Gamma_1  \rangle
\langle \Gamma_4 |  \hc_{\sigma'}\big )
\Big  \rangle_{\Psi_0}\;.
\end{eqnarray}
The  expressions for the on-site energy and the renormalisation matrix  
with a general
 orbital basis are given in Appendix~\ref{ka}.

\subsection{Energy functional}
In a translationally invariant system, the expectation values, which we 
introduced in the previous section, lead to the 
 following variational energy functional (per lattice site)
 \begin{eqnarray}\label{ap7.1}
E_{\rm G}\big (\lambda_{\Gamma,\Gamma'},\ket{\Psi_0}\big)&=&
\sum_{\substack{\sigma_1,\sigma_2 \\ \sigma'_1,\sigma'_2}}
q^{\sigma'_1}_{\sigma_1}\left(q^{\sigma'_2}_{\sigma_2}\right)^*
E_{\sigma_1,\sigma_2,\sigma'_1,\sigma'_2}\\\nonumber
&&+
\sum_{\Gamma,\Gamma_1,\Gamma_2}E_{\Gamma}
\lambda_{\Gamma,\Gamma_1}^{*}\lambda_{\Gamma,\Gamma_2}^{}
m^0_{\Gamma_1,\Gamma_2}\;.
\end{eqnarray} 
Here, we introduced the tensor 
\begin{eqnarray}\label{ap7.2}
E_{\sigma_1,\sigma_2,\sigma'_1,\sigma'_2}&\equiv&
\frac{1}{L}\sum_{i\neq j}   t^{\sigma_1,\sigma_2}_{i,j}
\langle  
\hcd_{i,\sigma'_1}\hc_{j,\sigma'_2}
\big \rangle_{\Psi_0}\\
&=& \frac{1}{L}\sum_{\ve{k}}
\varepsilon_{\ve{k};\sigma_1,\sigma_2}
\big \langle  
\hcd_{\ve{k},\sigma'_1}\hc_{\ve{k},\sigma'_2}
\big \rangle_{\Psi_0}
\end{eqnarray} 
with the bare dispersion
\begin{equation}
\varepsilon_{\ve{k};\sigma,\sigma'}\equiv\frac{1}{L}\sum_{i\neq j}
t^{\sigma,\sigma'}_{i,j}
e^{\rmi \ve{k}(\ve{R}_{i}-\ve{R}_{j}) }\;.
\end{equation} 
The energy~(\ref{ap7.1}) is a function of $\lambda_{\Gamma,\Gamma'}$
 and $\ket{\Psi_0}$ where $\ket{\Psi_0}$ enters~(\ref{ap7.1}), (\ref{ap7.2})  
 solely through the (non-interacting) density matrix $\tilde{\rho}$
with the elements 
\begin{equation}
\rho_{(i\sigma),(j\sigma')}\equiv
\langle \hat{c}_{j,\sigma'}^{\dagger}
\hat{c}_{i,\sigma}^{\phantom{+}}\rangle_{\Psi_0}\;.
\end{equation} 
Therefore, the energy 
\begin{equation}
E_{\rm G}=E_{\rm G}(\lambda_{\Gamma,\Gamma'},\tilde{\rho})
\end{equation} 
has to be minimised with respect to the 
 variational parameters $\lambda_{\Gamma,\Gamma'}$ and the 
density matrix $\tilde{\rho}$ obeying the 
constraints~(\ref{5.5}), (\ref{5.5b}), 
(or~(\ref{qwa}), (\ref{qwa2})) and 
\begin{equation}
\label{16} 
\tilde{\rho}^2=\tilde{\rho}\;.
\end{equation} 
This additional constraint ensures that $\tilde{\rho}$
corresponds to a single-particle wave function.

\section{Numerical Minimisation of the Gutzwiller Energy Functional}
\label{app7}

In principle, it is conceivable
 to minimise the energy with respect to the variational parameters 
$\lambda_{\Gamma,\Gamma'}$ and  the density matrix
$\tilde{\rho}$ simultaneously. However, we found it more efficient to 
  use consecutive cycles of `inner minimisations' (with respect to 
$\lambda_{\Gamma,\Gamma'}$ and with fixed $\tilde{\rho}$) and `outer 
minimisations' (with respect to $\tilde{\rho}$ and with fixed 
$\lambda_{\Gamma,\Gamma'}$) until a self-consistent minimum is reached.

In the following we assume that all quantities in the energy functional and
in the constraints are real. This is allowed since, in case of complex 
variational parameter or constraints~(\ref{5.5}), (\ref{5.5b}), we may 
 introduce the (independent) real and imaginary parts of these quantities. 

\subsection{`Inner' Minimisation}\label{app7.1}
Before we explain our minimisation algorithm in Sect.~\ref{inmin},  
it is essential to resolve the fundamental structure of our 
 energy function.
\subsubsection{Structure of the energy function}
 For a fixed density matrix $\tilde{\rho}$, the  energy function
 is given as
\begin{eqnarray}\nonumber
E_{\rm G}(\ve{v})&=&
\sum_{\sigma_1,\sigma_2,\sigma'_1,\sigma'_2}
q^{\sigma'_1}_{\sigma_1}(\ve{v})q^{\sigma'_2}_{\sigma_2}(\ve{v})
E_{\sigma_1,\sigma_2,\sigma'_1,\sigma'_2}\\\label{ap7.1dd}
&&+\sum_{Z,Z'}U_{Z,Z'}
v_Zv_{Z'}\;,
\end{eqnarray} 
where  we used the abbreviation 
$v_Z$ for the $n_{\rm v}$ variational 
parameters
\begin{equation}\label{ap7.1d}
v_Z=\frac{\lambda_{\Gamma,\Gamma'}}{\sqrt{m_{\Gamma}^0 m_{\Gamma'}^0}}\;,
\end{equation}
 which are 
considered as the elements 
 of a vector $\ve{v}$. 
In our numerical calculations 
we found that the inner minimisation, 
as it will be described in Sect.~\ref{inmin},
is much faster if we use the variational parameters
  (\ref{ap7.1d})  instead of 
 $\lambda_{\Gamma,\Gamma'}$.

  The renormalisation matrix
\begin{equation}\label{ap7.3}
q^{\sigma'}_{\sigma}(\ve{v})=\sum_{Z,Z'}S^{\sigma'}_{\sigma}(Z,Z')
v_Zv_{Z'}
\end{equation} 
and the $n_{\rm c}$ (independent) constraints~(\ref{5.5}), (\ref{5.5b}), 
which we denote as
\begin{equation}\label{ap7.7}
g_{l}(\ve{v})=\sum_{Z,Z'}f_{l}(Z,Z')v_Zv_{Z'}-g^0_{l}=0\;\;\;\;\;(l=1,\ldots,n_{\rm c})\;,
\end{equation} 
 are quadratic functions of the  variational parameters $v_Z$. The 
 numbers $g^0_{l}$ in~(\ref{ap7.7}) correspond to the r.h.s.\ 
of Eqs.~(\ref{5.5}), (\ref{5.5b}). 
 Note that, for a fixed density matrix $\tilde{\rho}$,  the coefficients
 $C_{Z,Z'}=\{S^{\sigma'}_{\sigma}(Z,Z'),f_{l}(Z,Z'),U_{Z,Z'}\}$ need 
 to be calculated only once. Moreover, we are free to work with an orbital 
 basis with a diagonal local density matrix, which allows us to calculate 
 these coefficients with the 
  simplified energy  expressions introduced in Sect.~\ref{chap2b}.
It is important in our algorithm that 
 the coefficients $C_{Z,Z'}$  are stored in the main memory of the 
computer because,
 in this way, derivatives of all quadratic functions can be 
 calculated very fast, see below. Even for large numbers $n_{\rm v}$  of 
 variational parameters  this can be achieved, since only a small fraction
 of the coefficients $C_{Z,Z'}$ is, in fact, finite and needs to be stored. 
In case that the main-storage capacity is exceeded, there are several 
 strategies to reduce the number of variational parameters, which we have 
 tested. They are discussed in Appendix~\ref{redpar}. 

The energy functional can be further simplified if we introduce
 the matrix 
\begin{equation}\label{ap7.5}
r^{\sigma'}_{\sigma}(\ve{v})
\equiv\sum_{Z,Z'}R^{\sigma'}_{\sigma}(Z,Z')
v_Zv_{Z'}
\end{equation} 
with the coefficients 
\begin{equation}\label{ap7.6}
R^{\sigma'_1}_{\sigma_1}\equiv 
\sum_{\sigma_2,\sigma'_2}E_{\sigma_1,\sigma_2,\sigma'_1,\sigma'_2}
S^{\sigma'_2}_{\sigma_2}(Z,Z')\;.
 \end{equation} 
It allows us to write the energy as
\begin{equation}\label{ap7.4}
E_{\rm G}(\ve{v})=
\sum_{\sigma_1,\sigma'_1}
q^{\sigma'_1}_{\sigma_1}(\ve{v})r^{\sigma'_1}_{\sigma_1}(\ve{v})+\sum_{Z,Z'}U_{Z,Z'}
v_Zv_{Z'}\;.
\end{equation}
Note that the coefficients in~(\ref{ap7.5}) also need to be calculated only 
 once in an inner minimisation and should be stored in the main memory.  
In this way, the 
 energy~(\ref{ap7.4}) and its gradient $\ve{E}(\ve{v})$ with the elements
\begin{eqnarray}\nonumber
E^{}_Z(\ve{v})&\equiv &\frac{\partial }{\partial v_{Z}}
E_{\rm G}(\ve{v})= 2\sum_{Z'}\Big[
\sum_{\sigma_1,\sigma'_1}
\big(q^{\sigma'_1}_{\sigma_1}(\ve{v})R^{\sigma'_1}_{\sigma_1}(Z,Z')\\\label{ap7.9}
&&+
r^{\sigma'_1}_{\sigma_1}(\ve{v})S^{\sigma'_1}_{\sigma_1}(Z,Z')\big)
+U_{Z,Z'}\Big]v_{Z'}
\;
\end{eqnarray}
can be calculated very fast. The same holds for the gradients 
$\ve{F}^{l}(\ve{v})$ of the constraints which have the elements 
\begin{equation}\label{ap7.8}
F^l_Z(\ve{v})\equiv \frac{\partial }{\partial v_{Z}}
g_{l}(\ve{v})= 2\sum_{Z'}f_l(Z,Z')
v_{Z'}\;.
\end{equation}
Note that in~(\ref{ap7.9}) and~(\ref{ap7.8}) we have used the symmetry
 $C_{Z,Z'}=C_{Z',Z}$, which we are free to impose.  
\subsubsection{Algorithm for the inner minimisation}\label{inmin}
We aim at a minimisation of the energy~(\ref{ap7.4}) in the 
manifold $\mathcal{M}_{\rm c}$ defined by the 
constraints~(\ref{ap7.7}). To this end, we can always start our minimisation
  in the uncorrelated limit, i.e., at the point
  $\ve{v}_0$ (with  
$\lambda_{\Gamma,\Gamma'}=\delta_{\Gamma,\Gamma'}$) for which
 $\ve{v}_0 \in \mathcal{M}_{\rm c}$ is 
 automatically fulfilled. We found numerical strategies that try  
  to move {\sl exactly}  along $\mathcal{M}_{\rm c}$ 
  to be quite cumbersome. Therefore, starting 
 from a certain point $\ve{v}_0\in \mathcal{M}_{\rm c}$,  we allow 
 the minimisation algorithm to violate the constraints
 by making `short' steps to 
 points $\ve{v}_1\notin \mathcal{M}_{\rm c}$. 
To keep the violation of the 
 constraints minimal, these steps have to take place in the   
 subspace $\mathcal{M}_{\parallel}(\ve{v}_0)$  
 that is tangential to $\mathcal{M}_{\rm c}$ at the point $\ve{v}_0$. 
 The optimal direction of a step in 
 $\mathcal{M}_{\parallel}(\ve{v}_0)$ is determined by 
 the tangential component of the gradient $\ve{E}(\ve{v}_0)$ since it 
  leads to a decrease of the  energy.       
In summary, and more precisely,  
these ideas lead to the following algorithm for  
 the inner minimisation:
\begin{itemize}
\item[i)\,] Find a point $\ve{v}_0$ in the variational parameter space $\mathcal{V}$ that 
 obeys the constraints~(\ref{ap7.7}) (i.e, $\ve{v}_0\in \mathcal{M}_{\rm c}$). 
\item[ii)\,] Determine the gradients  
$\ve{F}^{l}(\ve{v}_0)$ and $\ve{E}(\ve{v}_0)$. 
\item[iii)\,] Calculate the component $\ve{E}_{\parallel}(\ve{v}_0)$ 
of $\ve{E}(\ve{v}_0)$ in $\mathcal{M}_{\parallel}(\ve{v}_0)$ by 
 the following procedure.
The gradient $\ve{E}(\ve{v}_0)$  is written as 
\begin{equation} \label{ap7.10}  
  \ve{E}(\ve{v}_0)=\ve{E}_{\parallel}(\ve{v}_0)+
\ve{E}_{\perp}(\ve{v}_0)\;,
\end{equation}
where the tangential component $\ve{E}_{\parallel}(\ve{v}_0)$  is defined by 
\begin{equation}\label{ap7.11}
 \ve{E}_{\parallel}(\ve{v}_0)\cdot \ve{F}^{l}(\ve{v}_0)=0
\;\;\;\forall l\;.
\end{equation}  
The perpendicular component can be expressed as a linear combination 
\begin{equation}\label{ap7.12}
\ve{E}_{\perp}(\ve{v}_0)= \sum_{l=1}^{n_{\rm c}}\alpha_l 
\ve{F}^{l}(\ve{v}_0)
\end{equation}
 of the vectors 
$\ve{F}^{i}(\ve{v}_0)$.  In order to determine the 
coefficients $\alpha_i$, we multiply equation 
(\ref{ap7.10})
 with a vector $\ve{F}^{m}(\ve{v}_0)$ and use the expansion 
(\ref{ap7.12}). This leads to
 \begin{eqnarray}\label{ap7.13}
 \ve{E}(\ve{v}_0)\cdot \ve{F}^{m}(\ve{v}_0)
&=&\sum_{l}
\ve{F}^{l}(\ve{v}_0)\cdot\ve{F}^{m}(\ve{v}_0)\alpha_l\\\nonumber
&=&
\sum_{l}W_{m,l}(\ve{v}_0)\alpha_l\;,
  \end{eqnarray}
where we used equation~(\ref{ap7.11}) and introduced the (symmetric) 
matrix $\tilde{W}(\ve{v})$ with the elements 
\begin{equation}\label{ap7.14}
W_{m,l}(\ve{v})\equiv \ve{F}^{l}(\ve{v})\cdot\ve{F}^{m}(\ve{v})\;.
\end{equation}
The  linear equations~(\ref{ap7.13}) for  
 $\alpha_l$ have a unique solution, as
 long as the vectors $\ve{F}^{l}(\ve{v}_0)$ are linearly independent. 
A linear dependency of these vectors  can only arise if 
  certain constraints~(\ref{ap7.7}) are redundant. 
In that case, the redundant constraints 
 have to be eliminated right from the start.
With the coefficients  $\alpha_l$, we calculate the tangential component
\begin{equation}\label{ap7.16}
\ve{E}_{\parallel}(\ve{v}_0)=\ve{E}(\ve{v}_0)-
\sum_{l}\alpha_l\ve{F}^{l}(\ve{v}_0)\;.
\end{equation} 
of  $\ve{E}(\ve{v}_0)$.

\item[iv)\,] Make a `proper' step in the direction of   
$-\ve{E}_{\parallel}(\ve{v}_0)$ 
 to a new vector 
\begin{equation}\label{ap7.17}
\bar{\ve{v}}_1=\ve{v}_0-\beta \ve{E}_{\parallel}(\ve{v}_0)\;.
\end{equation}
 For the choice of the parameter $\beta$, various strategies are 
 conceivable. Since the point $\bar{\ve{v}}_1$ is not in  
$\mathcal{M}_{\rm c}$, the energy gain is not necessarily a useful 
 criterion and it is also rather time consuming to be determined. 
 Instead, we calculate 
 \begin{equation}\label{ap7.18}
 \Delta g(\bar{\ve{v}}_1) \equiv \sum_{l}[g_l(\bar{\ve{v}}_1)]^2\geq 0
 \end{equation}
as a measure for the violation of the constraints and choose the parameter 
  $\beta$ such that $\Delta g$ does not exceed a certain critical 
 value $\Delta g_{\rm c}$. This critical value should be 
automatically adjusted by  the algorithm 
 to ensure that, after returning to the hyper-surface $\mathcal{M}_{\rm c}$, 
there is a sufficient energy gain.
 \item[v)\,] In order to return to $\mathcal{M}_{\rm c}$
from the point $\bar{\ve{v}}_1\notin \mathcal{M}_{\rm c}$,  
 the following algorithm turned out to be very useful.
We seek a vector $\ve{v}_1$ that solves the constraint  equations 
 $g_l(\ve{v}_1)=0$ and is as close as possible to $\bar{\ve{v}}_1$.
 To this end, we could calculate  the gradients 
 $\ve{F}^{l}(\bar{\ve{v}}_1)$
 and try to solve the set of equations
\begin{equation}\label{ap7.19}
g_l\bigg(\bar{\ve{v}}_1+\sum_m \gamma_m\ve{F}^{m}(\bar{\ve{v}}_1)\bigg)=0
\end{equation}
by a proper choice of the coefficients $\gamma_m$. Such an  exact solution of
 equations~(\ref{ap7.19}), however, is quite time consuming. 
Therefore, we consider the linear set of equations 
\begin{equation}\label{ap7.20}
g_l(\bar{\ve{v}}_1)+\sum_{m}W_{l,m}(\bar{\ve{v}}_1)\gamma_m=0\;,
\end{equation}
which results from an expansion of~(\ref{ap7.19}) 
 to leading order in $\gamma_m$. Equations~(\ref{ap7.20}) can be readily
 solved with respect to $\gamma_m$.
This yields a new vector 
\begin{equation}\label{ap7.22}
\bar{\ve{v}}_1 \to\bar{\ve{v}}'_1=\bar{\ve{v}}_1+
\sum_m\gamma_m\ve{F}^{m}(\bar{\ve{v}}_1)\;.
\end{equation}
which, in general, is not yet a solution  of 
$g_l(\bar{\ve{v}}'_1)=0$. However, this vector is closer to 
$\mathcal{M}_{\rm c}$ than
$\bar{\ve{v}}_1$ because 
 $\Delta g(\bar{\ve{v}}'_1)<\Delta g(\bar{\ve{v}}_1)$.
 By an iteration of equations~(\ref{ap7.20})-(\ref{ap7.22})
  we eventually approach a vector   
$\ve{v}_1\in\mathcal{M}_{\rm c}$. Note that the fast convergence of this 
procedure is crucial for our algorithm. We have tried several other  ways 
 to return to $\mathcal{M}_{\rm c}$ that all turned out to be much slower. 
\item[vi)\,] If $E_{\rm G}(\ve{v}_1)<E_{\rm G}(\ve{v}_0)$
 we restart the procedure at point ii) with $\ve{v}_0$ replaced by  $\ve{v}_1$. 
 In case that $E_{\rm G}(\ve{v}_1)>E_{\rm G}(\ve{v}_0)$, 
the critical value $\Delta g_{\rm c}$ has to be lowered and the 
  algorithm  continues   
 with point iv). A useful measure for the convergence of the 
 whole iteration is the norm of $\ve{E}_{\parallel}$. This number 
 goes to zero near a 
 minimum $\ve{v}_{\rm min}$  of the energy functional 
$E_{\rm G}(\ve{v})$ for vectors $\ve{v}\in \mathcal{M}_{\rm c}$.    
\end{itemize}

\subsection{`Outer' Minimisation}\label{app7.2}
With the optimum variational parameters $\ve{v}^{\rm min}$ 
from the inner minimisation, described in Sect.~\ref{app7.1}, 
 we have to minimise the energy 
\begin{eqnarray}\label{4.7}
E_{\rm G}(\tilde{\rho})&=&\sum_{i\ne j}\sum_{\sigma,\sigma'}
\bar{t}^{\sigma,\sigma'}_{i,j}(\tilde{\rho})\rho_{(j\sigma'),(i\sigma)}\\\nonumber
&&+L\sum_{Z,Z'}U_{Z,Z'}(\tilde{\rho})
v^{\rm min}_Zv^{\rm min}_{Z'}
\end{eqnarray}
with respect to $\tilde{\rho}$. Here we introduced the renormalised 
hopping parameters
\begin{equation}\label{4.7b}
\bar{t}^{\sigma_1,\sigma_2}_{i,j}(\tilde{\rho})=\sum_{\sigma'_1,\sigma'_2}
q^{\sigma_1}_{\sigma'_1}(\tilde{\rho})q^{\sigma_2}_{\sigma'_2}(\tilde{\rho})
t^{\sigma'_1,\sigma'_2}_{i,j}
\end{equation}
and the renormalisation factors 
\begin{equation}\label{ap7.3sss}
q^{\sigma'}_{\sigma}(\tilde{\rho})=\sum_{Z,Z'}
S^{\sigma'}_{\sigma}(Z,Z';\tilde{\rho})
v^{\rm min}_Zv^{\rm min}_{Z'}\;.
\end{equation} 
In addition, the (independent) constraints~(\ref{qwa}), (\ref{qwa2}), 
\begin{eqnarray}\label{ap7.7b}
&&g_{l}(\tilde{\rho})=\sum_{Z,Z'}f_{l}(Z,Z',\tilde{\rho})
v^{\rm min}_Zv^{\rm min}_{Z'}-g^0_{l}=0\\\nonumber
&&(l=1,\ldots,n_{\rm c})\;,
\end{eqnarray}
and~(\ref{16})  need to be obeyed.

The local elements  of the density matrix 
\begin{equation}\label{782}
C_{\sigma,\sigma'} =\rho_{(i\sigma'),(i\sigma)}
\end{equation}
 play a special role in the 
 energy function because only they enter the coefficients in~(\ref{4.7}), 
  (\ref{ap7.3sss}), (\ref{ap7.7b}),
\begin{eqnarray}
U_{Z,Z'}(\tilde{\rho})&=&U_{Z,Z'}(\tilde{C})\;\;, \\
S^{\sigma'}_{\sigma}(Z,Z';\tilde{\rho})&=&S^{\sigma'}_{\sigma}(Z,Z';\tilde{C})\;\; , \;\;\\
f_{l}(Z,Z',\tilde{\rho})&=&f_{l}(Z,Z',\tilde{C})\;.
\end{eqnarray}
If they are kept fixed, only the hopping 
 term in~(\ref{4.7}) and the constraint~(\ref{16}) need to be taken  
 into account in the minimisation with respect to $\tilde{\rho}$. 
 This leads to a minimisation strategy which we discuss in Sect.~\ref{sw2}. 
An alternative way of minimising~(\ref{4.7}) with respect to {\sl all} 
 elements of $\tilde{\rho}$ will be introduced in  Sect.~\ref{sw1}.

The Hermiticity of the density matrix,
$\tilde{\rho}^{\dagger}=\tilde{\rho}^{}$, is a constraint which 
 is obeyed automatically in our outer minimisation algorithm  in 
Sect.~\ref{sw1}. To this end, however, the 
functional dependence of the energy with respect to $\tilde{\rho}^{}$, 
which is not unique, must be chosen such
 that 
\begin{equation}\label{axd}
\frac{\partial E_{\rm G}}{\partial \rho_{(i\sigma),(j\sigma')}}=
\left(\frac{\partial E_{\rm G}}{\partial \rho_{(j\sigma'),(i\sigma)}}\right)^*\;.
\end{equation} 
This can always be achieved by employing the Hermiticity of 
 $\tilde{\rho}$. We further assume that equation~(\ref{axd}) is also
 satisfied by the constraints~(\ref{ap7.7b}).

\subsubsection{Fixed local density matrix}\label{sw2}
If the local density matrix is fixed, we  have to minimise 
 \begin{equation}\label{4.7sss}
E_{{\rm G},0}(\tilde{\rho})\equiv\sum_{i\ne j}\sum_{\sigma,\sigma'}
\bar{t}^{\sigma,\sigma'}_{i,j}\rho_{(j\sigma'),(i\sigma)}
\end{equation}
with respect to $\tilde{\rho}$ obeying the constraints~(\ref{16}) 
and~(\ref{782}). We impose these constraints by means of 
Lagrange parameters $\eta_{\sigma,\sigma'}$ and $\Omega_{(i\sigma),(j\sigma')}$, 
 which leads to the `Lagrange functional'
\begin{eqnarray}\nonumber
L_{\rm G}&\equiv& E_{{\rm G},0}(\tilde{\rho})
-\sum_{\sigma,\sigma'}\eta_{\sigma,\sigma'}
\sum_i(C_{\sigma,\sigma'}-\rho_{(i\sigma'),(i\sigma)})\\\label{4.7bss}
&&-\sum_{i,j}\sum_{\sigma,\sigma'}\Omega_{(i\sigma),(j\sigma')}
[\tilde{\rho}^2-\tilde{\rho}]_{(j\sigma'),(i\sigma)}\;.
 \end{eqnarray}
As recalled in Appendix~\ref{ap3}, the minimisation of~(\ref{4.7bss}) with 
 respect to $\tilde{\rho}$ leads to 
 the effective single-particle Hamiltonian
\begin{equation}\label{tzs}
\hat{H}^{\rm eff}_0=\sum_{i\ne j}\sum_{\sigma,\sigma'}(\bar{t}^{\sigma,\sigma'}_{i,j}
+\delta_{i,j}\eta_{\sigma,\sigma'})\hcd_{i,\sigma}\hc_{j,\sigma'}\;.
\end{equation}
The optimum single-particle state $\ket{\Psi_0}$ is the ground state of 
 $\hat{H}^{\rm eff}_0$ where the parameters $\eta_{\sigma,\sigma'}$ have to be 
 chosen such that $C_{\sigma,\sigma'}=\langle \hcd_{i,\sigma}\hc_{i,\sigma'} \rangle_{\Psi_0}$ is satisfied.

With the state $\ket{\Psi}_0$, we may determine 
a new tensor~(\ref{ap7.2}) and start another run of the
 inner minimisation until self-consistency with respect to
  $\ket{\Psi}_0$ is reached. In this way, we find the ground-state energy 
$E=E_0(\tilde{C} )$ for a fixed local density matrix $C_{\sigma,\sigma'}$.
 To obtain the total variational ground-state energy,  $E_0(\tilde{C} )$
 still needs to be minimised with respect to $\tilde{C}$
 with the constraint of total particle number conservation, 
$\sum_{\sigma}C_{\sigma,\sigma}=N/L$. 
Alternatively, one may start a self-consistency cycle of inner 
and outer minimisation for a fixed set
of `effective crystal fields' $\eta_{\sigma,\sigma'} $ 
(and a fixed particle number). This defines an energy function  
$E_0(\tilde{\eta})$ which has to be minimised with respect to 
 $\eta_{\sigma,\sigma'}$.

Obviously, these two ways of minimising the energy are feasible only when
 the number $n_{\rm i}$ of independent elements in $\tilde{C}$ 
(or fields $\tilde{\eta}$) is small. 
It can also be useful, 
when there are physical reasons to minimise  $E_0(\tilde{C} )$
 (or $E_0(\tilde{\eta})$)
 only in some subspace of possible density matrices  $\tilde{C}$ (or fields 
 $\tilde{\eta}$)). 
Such a strategy has been used, e.g., in our calculations on the 
spin-orbit coupling effects in nickel. There, we could clearly identify
 the relevant fields 
$\eta_{\sigma}$: the dominant 
 term in nickel is the effective exchange splitting 
 accompanied by a smaller orbital-energy splitting and an effective
 spin-orbit coupling. In this way, the energy  $E_0(\tilde{\eta})$
 had to be minimised only in a $3$-dimensional subspace of 
 fields  $\tilde{\eta}$. However, such a procedure is bound to fail 
 when the number  $n_{\rm i}$ of parameters $\eta_{\sigma,\sigma'}$ is too 
large  and cannot be reduced by any physical 
arguments.  In that case, one may use the algorithm
 which we introduce in the following section.

\subsubsection{Unrestricted outer minimisation}\label{sw1}
In order to minimise the energy with respect to  
{\sl all} elements of the density matrix we impose the 
 constraints~(\ref{ap7.7b}) by means of Lagrange parameters $\Lambda_l$.
 This leads us to the functional 
\begin{eqnarray}\label{sgh}
L_{\rm G}&\equiv&
E_{\rm G}(\tilde{\rho})-\sum_l\Lambda_lg_{l}(\tilde{\rho})\\\nonumber
&&-\sum_{i,j}\sum_{\sigma,\sigma'}\Omega_{(i\sigma),(j\sigma')}
[\tilde{\rho}^2-\tilde{\rho}]_{(j\sigma'),(i\sigma)}
\end{eqnarray} 
where $E_{\rm G}(\tilde{\rho})$ has been defined in~(\ref{4.7}). The 
 minimisation with respect to $\rho$ yields again an
 effective single-particle Hamiltonian of the form~(\ref{tzs}) where
 the fields $\eta_{\sigma,\sigma'}$ are now given as 
\begin{equation}\label{sdfj}
\eta_{\sigma,\sigma'}=\frac{\partial}{\partial C_{\sigma,\sigma'}}E_{\rm G}(\tilde{\rho})-
\sum_l\Lambda_l\frac{\partial}{\partial C_{\sigma,\sigma'}}g_{l}(\tilde{\rho})\;.
\end{equation} 
To determine these fields we need to calculate the Lagrange 
parameters  $\Lambda_l$. This can by achieved if we use the fact that, in
 the variational ground state, the Lagrange functional (\ref{sgh})
 is also minimal 
 with respect to the variational parameters $v_Z$. This leads to
 the equations
 \begin{equation}
\frac{\partial}{\partial v_{Z}}E_{\rm G}(\tilde{\rho},\ve{v})\Big|_{\ve{v}
=\ve{v}^{\rm min}}-\sum_l\Lambda_l\frac{\partial}{\partial v_{Z}}
 g_{l}(\tilde{\rho},\ve{v})\Big|_{\ve{v}=\ve{v}^{\rm min}}=0
 \end{equation} 
which can be written in matrix-vector form as
\begin{equation}\label{dfg}
\tilde{G}\ve{\Lambda}=\ve{E}\;,
\end{equation} 
where $\tilde{G}$ and $\ve{E}$ have the elements
\begin{eqnarray}\label{dfg7}
\tilde{G}_{l,Z}&\equiv& \frac{\partial}{\partial v_{Z}}
 g_{l}(\tilde{\rho},\ve{v})\Big|_{\ve{v}=\ve{v}^{\rm min}}\;,\\
E_{Z}&\equiv&\frac{\partial}{\partial v_{Z}}
E_{\rm G}(\tilde{\rho},\ve{v})\Big|_{\ve{v}
=\ve{v}^{\rm min}}\;.
\end{eqnarray} 
The number of equations in~(\ref{dfg}) is usually much larger 
 then the number of parameters $\Lambda_l$. For physical reasons,
 however, Eq.~(\ref{dfg}) must have a unique solution. Therefore
 we can alternatively solve the equation
\begin{equation}\label{dfgd}
\tilde{G}^{\rm T}\tilde{G}\ve{\Lambda}=\tilde{G}^{\rm T}\ve{E}\;,
\end{equation} 
since it gives us the same solution for $\ve{\Lambda}$ as~(\ref{dfg}). 

 Note that the calculation of the 
 derivatives in~(\ref{sdfj}) is much easier if we work with an orbital 
 basis with a diagonal density matrix, see Appendix~\ref{ka4}. This leads
 us to the following algorithm for the outer minimisation.
\begin{itemize}
\item[i)\,] Set $q^{\sigma'}_{\sigma}=\delta_{\sigma,\sigma'}$ and choose a 
reasonable set of fields
$\eta^{({\rm i})}_{\sigma,\sigma'}$, e.g., 
$\eta^{({\rm i})}_{\sigma,\sigma'}=\varepsilon_{\sigma,\sigma'}$ 
 with the bare on-site energies $\varepsilon_{\sigma,\sigma'}$  
in the local Hamiltonian~(\ref{4.10a}).  
\item[ii)\,] Find the ground state $\ket{\Psi_0}$ of the effective 
Hamiltonian~(\ref{tzs}) with 
$\eta_{\sigma,\sigma'}=\eta^{({\rm i})}_{\sigma,\sigma}$
 and determine $C_{\sigma,\sigma'}$. If $C_{\sigma,\sigma'}$ is not 
 diagonal, 
find an orbital basis with a diagonal
 local density matrix. Continue the algorithm with this new basis 
  and its values for $C_{\sigma,\sigma'}=\delta_{\sigma,\sigma'}n_{\sigma}$
 and $E_{\sigma_1,\sigma_2,\sigma'_1,\sigma'_2}$. 
 \item[iii)\, ] Carry out an inner minimisation, as described in section 
\ref{app7.1}, and determine the Lagrange parameters $\Lambda_l$ by solving 
 Eq.~(\ref{dfgd}).
\item[iv)\,] Use Eq.~(\ref{sdfj}) to determine a new set of 
 parameters $\eta^{({\rm o})}_{\sigma,\sigma'}$. 
Set  $\eta^{({\rm i})}_{\sigma,\sigma'}\equiv \eta^{({\rm o})}_{\sigma,\sigma'}$ and
 go back to ii) until self-consistency, $\eta^{({\rm o})}_{\sigma,\sigma'}\approx 
\eta^{({\rm i})}_{\sigma,\sigma'} $ is reached.
\end{itemize}
This algorithm obviously relies on a certain `proximity' to the 
true variational
 ground-state, in particular, when there is more than one (local) minimum. 
 In the latter case, the algorithm 
 may have to be supported by a preliminary  manual scan of the 
variational space as 
described in Sect.~\ref{sw2}. Moreover, it can be necessary to introduce 
some kind of 'damping`  by setting
\begin{equation}
 \eta^{({\rm i})}_{\sigma,\sigma'}\equiv
\eta^{({\rm i})}_{\sigma,\sigma'}
+\beta
 (\eta^{({\rm o})}_{\sigma,\sigma'}-\eta^{({\rm i})}_{\sigma,\sigma'}) 
\end{equation} 
with $0<\beta<1$ instead of 
$\eta^{({\rm i})}_{\sigma,\sigma'}\equiv \eta^{({\rm o})}_{\sigma,\sigma'}$ 
in step iv). The value of $\beta$ must be small enough to ensure that 
 the energy decreases in each step of the cycle. In our numerical 
 tests, we found that $\beta$ may sometimes have to be smaller than $1$
 even in the immediate vicinity of the variational ground state. 

Note that the calculation of the 
derivatives in~(\ref{sdfj}) and~(\ref{dfg7}) in steps iii) and iv) of the 
algorithm is very much simplified by 
 the fact that the local density matrix is diagonal with respect to 
 $\ket{\Psi_0}$. This does {\sl not} mean, however, that  
 the derivatives with respect to non-diagonal elements $C_{\sigma,\sigma'}$  
 necessarily vanish, see Appendix\ref{ka4}. Therefore, the orbital 
 basis will, in general, be changing in each cycle of the algorithm until
 a self-consistent minimum is reached.   

\section{Summary}

In summary, we have given a detailed account of a numerical  
  scheme for the  minimisation of Gutzwiller energy functionals,
 which we found to be quite efficient in previous studies on 
 transition metals and transition metal compounds. We are confident 
 that our algorithm is of significant interest for other 
 researchers who intend to apply the multi-band Gutzwiller 
 theory to other materials.  

\begin{appendix} 

\section{Energy functional for an arbitrary local density matrix}
\label{ka}

The constraints~(\ref{5.5}), (\ref{5.5b}) for a general orbital basis 
read
\begin{eqnarray}\label{qwa}
\sum_{\Gamma,\Gamma_1,\Gamma_2}
\lambda_{\Gamma,\Gamma_1}^{*}\lambda_{\Gamma,\Gamma_2}^{}
m^{0}_{\Gamma_1,\Gamma_2}&=&1\;,\\\label{qwa2}
\sum_{\Gamma,\Gamma_1,\Gamma_2}
\lambda_{\Gamma,\Gamma_1}^{*}\lambda_{\Gamma,\Gamma_2}^{}
m^{0}_{\Gamma_1\cup \sigma,\Gamma_2\cup \sigma'}
&=&C_{\sigma,\sigma'}\;,
\end{eqnarray}
where  
\begin{eqnarray}
|\Gamma\cup \sigma \rangle
&\equiv& \hat{c}^{\dagger}_{\sigma}|\Gamma \rangle
 =\sum_{I(\sigma \notin I)}T_{I,\Gamma}|I \cup \sigma\rangle\;,\\\label{utr}
m^{0}_{\Gamma,\Gamma'}&=&\langle \hat{m}_{\Gamma,\Gamma'} \rangle_{\Psi_0}=
\sum_{I,I'}T_{I,\Gamma}T^*_{I',\Gamma'}m^{0}_{I,I'}\;,\\
m^{0}_{I,I'}&=&\langle \hat{m}_{I,I'}  \rangle_{\Psi_0}\;.
\end{eqnarray}
The result for the local energy is the same as in Eq.~(\ref{kdr}) only with 
 $m^{0}_{\Gamma_1,\Gamma_2}$ given by Eq.~(\ref{utr}). 

With Wick's theorem, the expectation values $m^{0}_{I,I'}$ in~(\ref{utr}) 
can be written as the determinant
\begin{equation}\label{miiprime}
m^{0}_{I,I'}=\left|
\begin{array}{cc}
\Omega^{I,I'}&-\Omega^{I,J}\\
\Omega^{J,I'}&\bar{\Omega}^{J,J}
\end{array}
\right|\;.
\end{equation}
Here, $\Omega_{I,I'}$ are the matrices
\begin{equation}
\Omega_{I,I'}=\left(
\begin{array}{cccc}
C_{\sigma_1,\sigma'_1}&C_{\sigma_1,\sigma'_2}&\ldots&C_{\sigma_1,\sigma'_{|I'|}}\\
C_{\sigma_2,\sigma'_1}&C_{\sigma_2,\sigma'_2}&\ldots&C_{\sigma_2,\sigma'_{|I'|}}\\
\ldots&\ldots&\ldots&\ldots\\
C_{\sigma_{|I|},\sigma'_1}&C_{\sigma_{|I|},\sigma'_2}&\ldots&C_{\sigma_{|I|},\sigma'_{|I'|}}
\end{array}
\right)\;,
\end{equation}
in which the entries are the elements of the 
uncorrelated local density matrix~(\ref{xc}), 
that belong to the configurations $I=(\sigma_1,\ldots,\sigma_{|I|})$ and
$I'=(\sigma'_1,\ldots,\sigma'_{|I'|})$. The matrix $\bar{\Omega}^{J,J}$ 
in~(\ref{miiprime}) is defined as 
\begin{equation}
\bar{\Omega}_{J,J}=\left(
\begin{array}{cccc}
1-C_{\sigma_1,\sigma_1}&-C_{\sigma_1,\sigma_2}&\ldots&-C_{\sigma_1,\sigma_{|J|}}\\
-C_{\sigma_2,\sigma_1}&1-C_{\sigma_2,\sigma_2}&\ldots&-C_{\sigma_2,\sigma_{|J|}}\\
\ldots&\ldots&\ldots&\ldots\\
-C_{\sigma_{|J|},\sigma_1}&-C_{\sigma_{|J|},\sigma_2}&\ldots&1-C_{\sigma_{|J|},\sigma_{|J|}}
\end{array}
\right)\;,
\end{equation}
with $\sigma_i\in J\equiv (1,\ldots,N)\backslash (I\cup I') $. 

The renormalisation matrix in~(\ref{8.410}) has the form 
\begin{eqnarray}\label{qmat}
q_{\sigma}^{\sigma'}&=&\sum_{\Gamma_1,\ldots,\Gamma_4}\lambda^{*}_{\Gamma_2,\Gamma_1}
\lambda^{}_{\Gamma_3,\Gamma_4}\langle \Gamma_2|\hat{c}^{\dagger}_{\sigma}  
|\Gamma_3 \rangle\\\nonumber
&&\times \sum_{I_1,I_4}T_{I_1,\Gamma_1}T^{*}_{I_4,\Gamma_4}
H^{\sigma'}_{I_1,I_4}\;,
\end{eqnarray}
where the matrix $H^{\sigma'}_{I_1,I_4}$ contains three different contributions 
depending on whether the index $\sigma'$ is an element 
 of $I_1\cap I_4$, 
$I_4\backslash (I_1\cap I_4)$, or $J=(1,\ldots,N)\backslash(I_1\cup I_4)$. 
With the abbreviation 
$f_{\sigma,I}\equiv\langle I |\hat{c}^{\dagger}_{\sigma}\hat{c}^{}_{\sigma} |I  \rangle$  
we can write $H^{\sigma'}_{I_1,I_4}$ as
\begin{eqnarray}\label{8sgdd}
H^{\sigma'}_{I_1,I_4}&\equiv&(1-f_{\sigma',I_1})\langle I_4  |\hat{c}^{}_{\sigma'} |I_4\cup \sigma'  \rangle
m^{0}_{I_1,I_4\cup \sigma'}\\\nonumber
&&+\left(
f_{\sigma',I_4}m^{0}_{I_1\backslash \sigma',I_4}+
(1-f_{\sigma',I_4})m^{0;\sigma'}_{I_1\backslash \sigma',I_4}
\right)\\\nonumber
&&\times \langle I_1 \backslash \sigma' |\hat{c}^{}_{\sigma'} |I_1  \rangle
\;.
\end{eqnarray}
The expectation value $m^{0;\sigma'}_{I_1\backslash \sigma',I_4}$ in~(\ref{8sgdd})
 has the same form as the one in~(\ref{miiprime}), except that the index $J$ 
 has to be replaced by $J \backslash \sigma'$. 

\section{Strategies to treat large numbers of `inner' variational parameters}
\label{redpar}

Our algorithm is particularly fast for the inner minimisation 
if we can store 
 all the second-order coefficients $C_{Z,Z'}$ in the main memory of our 
computer, see Sect.~\ref{app7.1}. Unfortunately, this cannot always be achieved
 in multi-band studies, in particular,
 when we include 
non-diagonal variational 
 parameters $\lambda_{\Gamma,\Gamma'}$. In this case we may try to reduce 
 the number of variational parameters, e.g., by symmetry considerations, 
see Appendix~\ref{redparb}. 
 Alternatively, one can employ additional numerical schemes that complement 
 our inner minimisation algorithm, see Appendix~\ref{redparc}.

\subsection{Reduction of the variational space }\label{redparb}
It is obvious that, due to symmetries, many parameters 
$\lambda_{\Gamma,\Gamma'}$ vanish automatically in the variational 
ground state and can be discarded from the outset. In order to identify
 these parameters one may use, e.g., the expectation values~(\ref{wet2}) 
which vanish for such parameters.

A further reduction can be achieved if we take only those 
variational parameters 
into account which couple states $|\Gamma\rangle,|\Gamma'\rangle$ that 
belong to the  same (degenerate) multiplet of the atomic Hamiltonian 
in~(\ref{4.10a}). Such a strategy has been used in our calculations
 on the spin-orbit coupling effects in nickel~\cite{buenemann2008}. 
 Although clearly an approximation, this scheme is justified since
 one is usually bound to make similar approximations already on 
 the level of the operators in the local Hamiltonian~(\ref{4.10a}). 
For example, in 
 studies on transition metals and their compounds  
 a spherical approximation is often used which allows one to express 
all Coulomb-interaction
 parameters by the three Racah or the three Slater--Condon parameters. To 
 go beyond this spherical approximation is actually simple within the 
Gutzwiller theory, however, it increases the number of independent 
Coulomb-interaction parameters significantly. Since there exists no 
established way to calculate these
 parameters from first principles, they have to be determined by 
 some fitting procedure, which only makes sense if their number is not too 
large.    

For sufficiently large Coulomb interactions, atomic charge fluctuations are 
 significantly suppressed. For example, in elementary nickel with its 
 approximately nine $3d$~electrons per atom the occupation of states with
 less than six $3d$-electrons is negligibly small. 
Hence, the variational parameters
$\lambda_{\Gamma,\Gamma'}$ of such shells may be assumed to be 
 diagonal or even to vanish.             
\subsection{Additional numerical schemes}\label{redparc}
In case that, even after all symmetry considerations, the number of 
variational parameters $\lambda_{\Gamma,\Gamma'}$ is still too large for our 
inner minimisation algorithm, one may employ one of the following numerical 
schemes.

The simplest scheme is to
split up the whole set of variational parameters into sub-sets, for 
 which the main storage of our computer is adequate and the minimisation 
algorithm in 
Sect.~\ref{inmin} can  be applied. The minimisation with respect to each of 
these  sub-sets of parameters has then to be repeated until 
 a total minimum is reached. 
 
Another  scheme is based on the observation that the multiplet states 
$|\Gamma\rangle$ do not necessarily have to be the eigenstates of 
 our local Hamiltonian~(\ref{4.10a}). Instead, the states $|\Gamma\rangle$
 themselves are  considered as variational objects in the following
 algorithm.
\begin{itemize}
\item[(i)\,] Choose a certain basis of multiplets states 
$|\Gamma\rangle^{\rm (i)}$
\item[(ii)\,] Set $|\Gamma\rangle=|\Gamma\rangle^{\rm (i)}$ and determine the 
most `relevant' non-diagonal variational parameters
 $\lambda_{\Gamma,\Gamma'}$ such that their number still allows for the use
 of the minimisation algorithm in Sect.~\ref{inmin}. A criterion for the 
 `relevance' of the parameters $\lambda_{\Gamma,\Gamma'}$ may be the size 
 of the non-interacting expectation value~(\ref{wet2}). Alternatively one 
 could use the corresponding correlated expectation value which can 
be calculated in a preceding calculation with a diagonal variational parameter matrix
 $\lambda_{\Gamma,\Gamma}$.  
 \item[(iii)\,] Determine the optimum values $\lambda^{\rm opt}_{\Gamma,\Gamma'}$ of the 
parameters chosen in (ii). Calculate the eigenstates   
$|\Gamma\rangle^{\rm (o)}$ of the optimal correlation operator 
\begin{equation}
\hat{P}^{\rm opt}=\sum_{\Gamma,\Gamma'}\lambda^{\rm opt}_{\Gamma,\Gamma'}
\hat{m}_{\Gamma,\Gamma'}\;.
\end{equation}
\item[(iv)\,] Set $|\Gamma\rangle^{\rm (i)}=|\Gamma\rangle^{\rm (o)}$ and 
 go back to (ii) until self-consistency 
$|\Gamma\rangle^{\rm (i)}\approx|\Gamma\rangle^{\rm (o)}$ is 
reached.
\end{itemize}   
We have tested both numerical schemes, discussed in this Appendix. 
 From these preliminary calculations, however, we are not yet able
  to draw any final conclusions on the efficiency of both approaches.   

\section{Minimisation of functions with respect to non-interacting 
density matrices}\label{ap3}

We consider a general function $E(\tilde{\rho})$ of a non-interacting 
 density matrix $\tilde{\rho}$ with the elements
 \begin{equation}
\rho_{\gamma,\gamma'}=\langle \hcd_{\gamma'} \hc_{\gamma}\rangle_{\Phi_0}\;.
\end{equation}
The fact that  $\tilde{\rho}$ is derived from a 
 single-particle product  wave function $\ket{\Phi_0}$ is equivalent to the 
 matrix equation $\tilde{\rho}^2=\tilde{\rho}$. 
Hence, the minimum of $E(\tilde{\rho})$ in the `space' of all  
{\sl non-interacting} 
 density matrices is determined by the condition
\begin{equation}
\frac{\partial}{\partial \rho_{\gamma',\gamma}}L(\tilde{\rho})=0\;,
\end{equation}
where we introduced the `Lagrange functional'
\begin{eqnarray}\label{sfg}
L(\tilde{\rho})&\equiv& E(\tilde{\rho})-\sum_{l,m}\Omega_{l,m}
\big[\tilde{\rho}^2-\tilde{\rho}\big]_{m,l} \\
&=& E(\tilde{\rho})-\sum_{l,m}\Omega_{l,m}\Big(
\sum_p \rho_{m,p} \rho_{p,l}-\rho_{m,l}\Big)
\end{eqnarray}
and the matrix $\tilde{\Omega}$ of Lagrange parameters $\Omega_{l,m}$. 
The minimisation of~(\ref{sfg}) leads to the matrix equation
\begin{equation}
\tilde{H}=\tilde{\rho}\tilde{\Omega}+\tilde{\Omega}\tilde{\rho}-\tilde{\Omega}
\end{equation}
for the `Hamilton matrix' $\tilde{H}$ with the elements 
\begin{equation}
H_{\gamma,\gamma'}=
\frac{\partial}{\partial \rho_{\gamma',\gamma}}
 E(\tilde{\rho})\;.
\end{equation}
This equation is satisfied if $\tilde{\rho}^2=\tilde{\rho}$ and
\begin{equation}
[\tilde{H},\tilde{\rho}]=0\;.
\end{equation}
Hence, $\tilde{H}$ and $\tilde{\rho}$ must have the same basis 
 of (single-particle) eigenvectors and, consequently, 
$\ket{\Phi_0}$ is the ground state of
 \begin{equation}
\hat{H}_0^{\rm eff}=\sum_{\gamma,\gamma'}H_{\gamma,\gamma'}
\hcd_{\gamma} \hc_{\gamma'}\;.
\end{equation}

\section{Derivatives of the general energy functional} \label{ka4}

In Sect.~\ref{sw1}, we have to calculate the derivative of the ground-state 
 energy and of the constraints with respect to the elements of the
 local density matrix, see Eq.~(\ref{sdfj}). 
Equations~(\ref{qwa})--(\ref{8sgdd})
 reveal that, in fact, we only need the derivatives of  
 $m^0_{I,I'}$ (and of $m^{0;\bar{\sigma}}_{I\backslash \bar{\sigma} ,I'}$). 
For a general density matrix 
$C_{\sigma,\sigma'}$, their calculation requires an evaluation of 
determinants such as~(\ref{miiprime}). However, in Sect.~\ref{sw1} we work
 with an orbital basis for which $C_{\sigma,\sigma'}
=\delta_{\sigma,\sigma'}n_{\sigma}$. Hence the derivatives with respect to
 $C_{\sigma,\sigma'}$ have a much simpler form. For example, for the 
derivatives of 
 $m^0_{I,I'}$ we find
\begin{eqnarray}
\frac{\partial }{\partial C_{\sigma,\sigma}}m^0_{I,I'}=\delta_{I,I'}m^0_{I,I}
\left\{ 
\begin{array}{cl}
1/n_{\sigma}&{\rm for}\;\;  \sigma \in I\\
-1/(1-n_{\sigma})&{\rm for}\;\; \sigma \notin I
\end{array}
 \right.
\end{eqnarray}
for $\sigma=\sigma'$, and
\begin{equation}
\frac{\partial }{\partial C_{\sigma,\sigma'}}m^0_{I,I'}
=\delta_{\bar{I},I \backslash \sigma}
\delta_{\bar{I},I'\backslash \sigma'}\frac{m^0_{\bar{I},\bar{I}}}
{(1-n_{\sigma})(1-n_{\sigma'})}
\end{equation}
for $\sigma\ne \sigma'$, where $\sigma \in I$ and $\sigma' \in I'$. The 
 derivatives of $m^{0;\bar{\sigma}}_{I\backslash \bar{\sigma} ,I'}$ are 
 given accordingly.

 \end{appendix} 

\providecommand{\WileyBibTextsc}{}
\let\textsc\WileyBibTextsc
\providecommand{\othercit}{}
\providecommand{\jr}[1]{#1}
\providecommand{\etal}{~et~al.}

\end{document}